\documentclass{article}
\usepackage{spconf,amsmath,graphicx}

\usepackage{amsmath,graphicx}
\usepackage{multirow}
\usepackage[hang,flushmargin]{footmisc}
\usepackage{caption} 
\usepackage[tableposition=top]{caption}
\usepackage{hyperref,color,soul}
\hypersetup{colorlinks=true}

\title{Cleanformer: A multichannel array configuration-invariant neural enhancement frontend for ASR in Smart Speakers}

\name{Joseph Caroselli, Arun Narayanan, Nathan Howard, Tom O'Malley}
\address{Google LLC, U.S.A.}

\begin{document}
\ninept
\maketitle
\begin{abstract}
This work introduces \emph{Cleanformer} \textemdash a streaming multichannel neural enhancement frontend for automatic speech recognition (ASR). This model has a Conformer-based architecture which takes as inputs a single channel each of raw and enhanced signals, and uses self-attention to derive a time-frequency mask. The enhanced input is generated by a multichannel adaptive noise cancellation algorithm known as Speech Cleaner. The time-frequency mask is applied to the noisy input to produce enhanced features for ASR. Detailed evaluations are presented with speech- and non-speech-based noise that show significant reduction in word error rate (WER) -- about 80\% for -6~dB SNR -- over a state-of-the-art ASR model alone. It also significantly outperforms enhancement using a beamformer with ideal steering. The enhancement model can be used with different microphone arrays without the need for retraining.

%Abstract is expected to be <1000 characters.  Below is the original.
% This work introduces the \emph{Cleanformer}, a streaming multichannel neural enhancement frontend for automatic speech recognition (ASR). This model takes as inputs a single channel each of raw and enhanced signals, and uses a conformer-based architecture to derive a time-frequency mask. The enhanced input is generated by a multichannel adaptive noise cancellation algorithm known as Speech Cleaner, which makes use of noise context to derive its filter taps. The time-frequency mask is applied to the noisy input to produce enhanced output features for ASR. Detailed evaluations are presented with simulated and re-recorded datasets in speech-based and non-speech-based noise that show significant reduction in word error rate (WER) when using a large-scale state-of-the-art ASR model.  It also will be shown to significantly outperform enhancement using a beamformer with ideal steering.  The enhancement model is agnostic of the number of microphones and array configuration and, therefore, can be used with different microphone arrays without the need for retraining. It is demonstrated that performance improves with more microphones, up to 4, with each additional microphone providing a smaller, but significant, benefit. Specifically, for an SNR of -6\,dB, relative WER improvements of about 80\% are shown in speech and non-speech noisy conditions over a state-of-the-art ASR model alone.
\end{abstract}
\vspace{-0.05in}
\section{Introduction}
\label{sec:introduction}
%Robustness in ASR

Robustness of automatic speech recognition in the presence of noise has taken significant strides in recent years due to the adoption of neural network based acoustic models \cite{PrabhavalkarRaoSainathLiEtAl17,BattenbergChenChildCoatesEtAl17,li2021betterfaster}, large scale training \cite{mirsamadi2017multi,hakkani2016multi,NarayananMisraSimPundakEtAl18}, and improved data augmentation strategies \cite{kim2017mtr, park2019specaugment, medennikov2018investigation}. However, conditions like reverberation, loud background noise, and competing speech still pose  challenges for ASR models \cite{barker2018fifthchime}. Consequently, speech enhancement frontends for ASR that specifically address background noise have been widely studied \cite{Li2014Review, zhang2018deep}. 

%Unique challenges for smart speakers.

Improving ASR performance on smart speakers is the focus of this work.
Smart speakers are a widespread application of ASR today and operate in noisy household environments that include kitchen fans, television and screaming children. 
These devices present some specific characteristics. The queries directed at them are typically just a few seconds long and prefaced by a keyword. The user expects a speedy response meaning latency is of paramount importance. When there are multiple speakers or speech-based noise, there is only one desired speaker to whom the device should respond \textemdash the person who spoke the keyword. Smart speakers possess some advantages that can help them deal with these challenges. Commonly available is an array of two or more microphones that can be used for spatial processing. On-device processing can be performed using pre-utterance audio to better understand the noise context. 

Combining neural modeling with signal processing algorithms is common in enhancement. One popular technique is a beamformer \cite{benesty2008microphone} which is steered using statistics of a desired source estimated using a time-frequency mask generated by a neural network \cite{higuchi2016robust,heymann2016neural}. This can be very effective when the desired speaker is in the presence of non-speech noise but less so when deciding between one or more voices. Moreover, these techniques failed to demonstrate the same efficacy when operating under the streaming low-latency constraints of smart speakers \cite{heymann2018performance,li2016neural}. Other solutions are aimed at separating multiple speakers using a single microphone \cite{hershey2016deep,luo2019conv,luo2020dual} or utilizing the spatial information provided by a microphone array \cite{chang2019mimo,chang2020end}. However, these are often designed for separating multiple voices rather than identifying one target. Techniques that target a particular speaker often make use of speaker-id \cite{wang2018voicefilter} or noise context \cite{huang2019hotword}.

% Recent work has shown promising results with signals enhanced by signal processing techniques used as input to enhancement models. For example, \cite{liu2022neural} and \cite{wang2021hybrid} use multiple beamformed signals as input to a neural network while \cite{lu2022towards} iterates between a beamformer and a network.
%Proposed solutions
% A particular challenge is the multi-talker scenario, where more than one person is speaking. This is especially true for smart speakers where it is desired to respond to one of the speech sources and not the others, whether they are from television, radio, or other people. In such a scenario, the desired speaker needs to be determined and isolated from the other sources. There have been several proposed solutions aimed at separating the multiple speakers using a single microphone \cite{hershey2016deep,luo2019conv,luo2020dual} or taking advantage of the spatial information provided by a microphone array \cite{chang2019mimo,chang2020end}. However, these are often designed for separating multiple voices rather than identifying one target. Techniques that target a particular speaker often make use of speaker-id \cite{wang2018voicefilter} or the noise context \cite{huang2019hotword}.

\begin{figure}[t]
 \centering
 \includegraphics[scale=0.3]{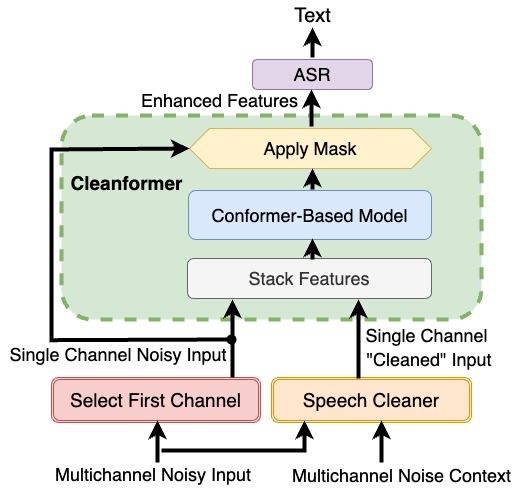}
  \vspace{-0.1in}
 \caption{Cleanformer architecture}
  \label{fig:cleanformer}
  \vspace{-0.3in}
\end{figure}

% A signal processing technique that has been applied successfully in a streaming setting is Hotword Cleaner \cite{huang2019hotword}. This adaptive noise cancellation algorithm has shown significant improvement in hotword recognition in noisy environments. In this work, the underlying algorithm is re-purposed to process query speech and will be referred to as \emph{Speech Cleaner}. The Speech Cleaner utilizes the audio directly before the hotword or wake-word (``Alexa" or ``OK Google"), referred to as the noise context. It is able to do this because the algorithm operates on device, and continuously adapts its filter taps. Enhancement, at each time-step, is performed using filter coefficients buffered during adaptation from a few hundred milliseconds in the past, which ensures that speech from the desired speaker (the one speaking the wake-word) is not used to learn the filter taps.

This work introduces \emph{Cleanformer}, a Conformer \cite{gulati2020conformer}-based multichannel neural enhancement frontend for ASR. Promising results have been shown in recent works using one or more signals enhanced by signal processing techniques, usually beamformers, as input to enhancement models \cite{liu2022neural,wang2021hybrid,wang2022stft}. In a similar vein, the inputs to Cleanformer are a single channel of raw (unprocessed) audio and a single channel of enhanced output. The enhanced output comes from the \emph{Speech Cleaner}, a multi-channel adaptive noise cancellation algorithm that has shown significant improvement in keyword recognition robustness \cite{huang2019hotword, Huang2019} and is applied here to process query speech. Together these input signals are used to estimate a time-frequency mask. The mask is applied to the noisy input features to produce estimates of the clean input log-mel features. These features serve as the input to a separately trained ASR model.  The overall architecture of Cleanformer is shown in Figure \ref{fig:cleanformer}.  

While it is designed to operate with a multi-channel array, Cleanformer itself is agnostic to the number of microphones in the array or their layout and, therefore, does not require re-training for different arrays. As the array size changes, the only difference is the number of input channels that Speech Cleaner receives. 

Our contributions include: 1) targeting one desired speaker in a multitalker environment without speaker-id, 2) model reuse without retraining for different arrays, 3) a signal enhanced through adaptive noise cancellation as an input to a model along with the raw unenhanced signal and 4) significant performance improvements under the streaming low-latency requirements of smart speakers.

We will demonstrate that Cleanformer reduces WER, often by greater than 50\%, in noisy environments, whether speech- or non-speech based. Increasing the number of microphones is shown to improve performance with each additional microphone providing smaller, but significant, gains. The rest of the paper is organized as follows:  the Cleanformer enhancement model is presented in {Section~\ref{sec:model}}, the experimental setup is described in {Section~\ref{sec:exp}}, {Section~\ref{sec:results}} details the results and conclusions are presented in {Section~\ref{sec:conclusion}}.

\vspace{-0.05in}

\section{Cleanformer}
\label{sec:model}
% The overall architecture of Cleanformer is shown in Figure \ref{fig:cleanformer}. It takes a single channel each of raw noisy features and enhanced features as inputs, and estimates a time-frequency mask to filter out unwanted signals. The mask is applied to the noisy input features to produce estimates of the clean input log-mel features. These features serve as the input to an ASR model. 
\vspace{-0.05in}
\subsection{Speech Cleaner}

\label{sec:cleaner}

The enhanced input features used in this model are generated using Speech Cleaner, an adaptive noise cancellation algorithm. The algorithm is described in depth in  \cite{huang2019hotword, Huang2019} where it was been applied to keyword detection. Here, it is applied directly to the target query in addition to the keyword. As shown in Figure \ref{fig:cleanformer}, Speech Cleaner utilizes the multichannel noise context (the audio before the keyword is detected) to estimate the filter coefficients, which are then applied to the multichannel noisy input. The functionality and implementation are very similar to a standard linear acoustic echo cancellation filter \cite{tashev2009sound}, with a modified input configuration. 

% Because Speech Cleaner functions on device, it can make use of the signal that occurs directly before the keyword which serves as the noise context. As the desired speaker is expected to be the person speaking the keyword, we can assume with high confidence that the desired speaker is present during this time segment. In the period of time directly before the keyword, it is assumed that the desired speaker is not speaking. Therefore, statistics of the noise can be estimated from the noise context, and used for noise reduction during the keyword and the subsequent query.

Speech Cleaner operates on multichannel STFT-processed input signals independently for each frequency to produce a single channel of enhanced STFT-domain output:

\vspace{-0.1in}
\newcommand{\UM}{\mathbf{U}_{m}(k)}
\newcommand{\UMOPT}{\mathbf{U}^{*}_m(k)}
\begin{equation}
\label{eq:filter}
    Z(n,k)=Y_{0}(n,k)-\sum_{m=1}^{M-1}[{\UM^H}{\mathbf{Y}}_{m}(n,k)].
\end{equation}
 $\mathbf{Y}_m(n,k) = [Y_m(n,k), \cdots, Y_m(n-(L-1),k)]^T$, is a vector of time-delayed STFT-processed input corresponding to frames $n$ through $n-(L-1)$ for microphone $m$ and frequency $k$. In all examples presented here $L$ is set to 3. $(\cdot)^H$ represents the Hermitian operator.  To the input from each of the $M$ microphones except the first ($m=0$), a finite impulse response (FIR) filter with coefficients $\UM$ is applied. The summed output of these $M-1$ filters is subtracted from frame $n$ of the first microphone, $Y_{0}(n,k)$ to produce the enhanced output, $Z(n,k)$.
 
%  Finite impulse response(FIR) filters of length $L$ are applied to the signals from all microphones except one arbitrarily selected here to be the index 0 microphone . The summed output of these filters is subtracted from the index 0 microphone to produce   Mathematically,
% $\mathbf{Y}_m(k,n)$ is a vector of time-delayed STFT-processed input corresponding to frames $n$ through $n-(L-1)$ for microphone $m$ and frequency $k$. and $\mathbf{U}_m(k,n)$ is a vector of the $L$  coefficients used to fiter that vector. $M$ is the number of microphones. In all examples presented in this work $L$ is set to 3.  These coefficients are dependent on the frame number because 
The optimal filter for cancelling the noise during the keyword and the query, $\UMOPT$, is estimated by minimizing the expectation of output power over all frames during the \emph{noise context} right before the keyword, when it is assumed that there is no desired speech present:
\begin{equation}
\UMOPT=\underset{\UM}{\mathrm{argmin}}~E_n{[|Z(n,k)|^2]}.
\end{equation}
In practice, the filters are learned via continuous adaptation using the recursive least squares (RLS) algorithm \cite{Haykin:2002}. A forgetting factor is used so that only the previous few seconds impact the coefficient values at any point in time. When the keyword is detected, the adaptation is stopped and buffered filter coefficients from a few hundred milliseconds in the past (before the keyword) are applied to the keyword and the query to produce the enhanced output. This ensures that the filter is adapted before the desired speaker starts speaking, while enabling the filter to cancel the noise right before the keyword. We assume that the spatial profile of the noise does not change significantly for the duration of the noise context, the keyword and the query, which is usually true in a smart speaker setting (e.g., a TV playing continuously in the background). 

% Adaptation in time is performed using recursive least squares (RLS) \cite{Haykin:2002} with the objective of obtaining filter coefficients that minimize the expectation of the output power over all frames:
% \newcommand{\argminD}{\arg\!\min} % AlfC
% \begin{equation}
%     \hat{\mathbf{U}}_m(k)=\argminD_{\mathbf{U}_m(k,n)}  E_n{[|Z(k,n)|^2]}.
% \end{equation}
%  The adaptation occurs before the keyword is detected(during the noise context) when it is assumed there is no desired speech present. While the adaptation is ongoing until keyword detection, it is performed with a forgetting factor such that only the previous few seconds impact the coefficient values. When the keyword is detected, the adaptation is stopped and buffered filter coefficients from a few hundred milliseconds in the past (before the keyword) are then applied to the query to produce the enhanced output. This ensures that the filter adapts before the desired speaker was present and enables the filter to cancel the noise but not the desired signal because the desired speaker was not present during adaptation of the cancellation filter.

\vspace{-0.1in}
\subsection{Conformer}
\vspace{-0.05in}

Cleanformer is based on the Conformer architecture \cite{gulati2020conformer}. We use the open-sourced implementation \cite{lingvo} of the Conformer layer \cite{li2021betterfaster}, which consists of a half-step feed-forward module, a convolution module, a multi-head self-attention module and another half-step feed-forward module. The convolutional block is comprised of point-wise convolution, gated linear units, 1-D depth-wise convolution, and group normalization. Residual connections are present between each block. Layer normalization takes place before each processing block as well as after the final half-step feed-forward module. We use Conformers as they are especially well suited for streaming applications since they are more easily parallelizable than LSTMs, as shown in prior works in ASR \cite{li2021betterfaster} and separation \cite{chen2021continuous}.

\vspace{-0.1in}
\subsection{Implementation Details}
% \vspace{-0.05in}
% \subsubsection{Features}
% \vspace{-0.1in}

The enhancement frontend takes as input one channel of raw audio and the single channel Speech Cleaner output. Each input is converted to the 128-dimensional log-mel domain using a 32~ms window with 10~ms step. Four frames from each of the two sources are stacked at the input, and are then subsampled by a factor of 3, resulting in a feature representation with a 30~ms step. This matches the representation used by the ASR model.

% \vspace{-0.15in}
% \subsubsection{Target}
% \vspace{-0.1in}

For the training target, the ideal ratio mask (IRM) \cite{Narayanan2013IRM} is used. It is computed in the mel-spectral space using reverberant speech and reverberant noise, with the assumption that they are uncorrelated:
\begin{equation}
\label{eq:ratio}
M(n,f) = \frac{{\mathbf X}(n,f)}{{\mathbf X}(n,f) + {\mathbf N}(n,f)}. 
\end{equation}
$\mathbf{X}$ and $\mathbf{N}$ represent, respectively, mel filterbank magnitudes of the reverberant speech and reverberant noise and $f$ is the mel channel. Using the IRM as the target enables enhancement to be performed directly in the feature space eliminating the need to reconstruct the waveform.

% \vspace{-0.15in}
% \subsubsection{Loss}
% \vspace{-0.05in}

A combination of two losses is used during training as recommended in \cite{howard2021neural}. The first is a spectral loss that is a combination of  $\ell_1$ and $\ell_2$ losses between the IRM and the estimated IRM $\widehat{M}(n,f)$: 
\begin{equation}
\label{eq:irm_loss}
\mathcal{L} = \sum_{n,f} (|M(n,f) - \widehat{M}(n,f)| + (M(n,f) - \widehat{M}(n,f))^2).
\end{equation}
The second is an ASR-based loss. It is computed by passing log filterbank energies of the target utterance and those produced by the enhancement frontend to a pre-trained end-to-end ASR model. The loss is computed using only the ASR model encoder \cite{howard2021neural}. The $\ell_2$ distance between the encoder output of the target features and that of the enhanced features is minimized. The ASR model encoder is kept fixed during training to decouple it from the enhancement model.

% \vspace{-0.15in}
% \subsubsection{Inference}
% \vspace{-0.05in}

During inference, the estimated IRM is scaled and floored. This will reduce the amount of speech distortion in the masked output at the expense of diminished noise suppression. The enhanced estimate of the clean mel spectrogram $\mathbf{\widehat{X}}$ is obtained by applying the scaled and floored estimated mask to the noisy mel spectrogram $\mathbf{Y}$ via pointwise multiplication:
\begin{equation}
\label{eq:masking}
\mathbf{\widehat{X}}(n,f) = \mathbf{Y}(n,f) \odot \max(\widehat{M}(n,f)^\alpha, \beta).
\end{equation}
$\alpha$ and $\beta$ are the exponential mask scalar and mask floor, respectively. A neural network is used to perform mask scalar selection on a frame-by-frame basis \cite{narayanan2022mask}, with $\beta=0.01$ chosen after tuning on development sets. The output is log compressed and ${\log{\mathbf{\widehat{X}}}}$ is passed to the ASR model which has been trained separately. 

% Because the ASR model is sensitive to speech distortion and non-linear processing, this can have an impact on performance \cite{Narayanan2014Joint}.
% In all experiments, $\alpha$ is set to 0.5 and $\beta$ to 0.01, after tuning on development sets. The output is log compressed, i.e., ${\log{\mathbf{\widehat{X}}}}$, and passed to the ASR model.

% \vspace{-0.15in}
% \subsubsection{Model Architecture}
% \vspace{-0.05in}

The enhancement frontend consists of 4 Conformer layers each having 256 units. The feed-forward module has 1024 dimensions and the kernel size in the convolution module is 15. The self-attention modules apply masked attention with 8 heads. Each frame attends to 31 frames in the past. Only past frames are used so as to enable a streaming model. After the final Conformer layer, a single fully-connected layer with sigmoid activation is utilized. The model has approximately 6.5M parameters.

% \vspace{-0.15in}
% \subsubsection{ASR Model}
% \vspace{-0.05in}

A recurrent neural transducer model with LSTM-based encoder layers \cite{sainath2020streaming} is used for ASR evaluations. This model was pre-trained independently of Cleanformer using approximately 400k hours of anonymized, hand-transcribed English utterances from domains like Search, Telephony, and YouTube. Data augmentation has been applied during training to simulate SNR values from 0 to {$30$~dB} and reverberation with $T_{60}$ from 0 to 900~ms. SpecAug \cite{park2019specaugment} is also used. This model takes as input log-mel features of the same characteristics as those produced by Cleanformer.

\vspace{-0.05in}

\section{Experimental Settings}
\label{sec:exp}
\vspace{-0.1in}
\subsection{Training}
\label{sec:training}
\vspace{-0.05in}

For training, $\sim$50k hours of speech are used based on Librispeech \cite{panayotov2015LibriSpeech}, Librivox \cite{kearns2014librivox} and internal vendor-collected utterances. There are 281k utterances in Librispeech and 1.9M in the vendor-collected set. Librivox is segmented to create utterances that are 3 to 15 seconds long, resulting in a dataset with 18M utterances. A room simulator \cite{kim2017mtr} is used to add reverberation and noise to these utterances and to model reception by a 3-microphone triangular array. {$T_{60}$} times range from 0 ms to 900 ms are used while SNR is between {$-10$} and {$30$~dB}. Noise is taken from internally collected sets in conditions like cafes, kitchens and cars, freely available noise sources Getty \cite{getty} and YouTube Audio Library \cite{youtube-audio}, and, to simulate multi-talker conditions, randomly selected speech from the training sets. 

Each query is prefaced with roughly 6 seconds of noise to serve as the noise context. This length is arbitrary as this context is used to emulate the continual adaptation of the Speech Cleaner prior to keyword detection. As stated previously, only the previous few seconds impact the coefficient values. In order to make Cleanformer more robust to the failure of meeting the assumption that the desired speaker is not present in the noise context, the noise context is replaced in 20\% of the training utterances. In half of those cases, the context was replaced with the query itself to ensure that the desired speaker is present. In the other half, white noise was used for the noise context which prevents Speech Cleaner from learning anything useful to remove noise in the query portion of the input audio. Note that the noise context is only used by Speech Cleaner to adapt the filter taps and is never passed to Cleanformer or the ASR model.

% Each of the original utterances is used to generate multiple noisy utterances with different noise and room configurations to increase the diversity of the training set. 

% \footnote{\url{https://www.gettyimages.com/about-music}}
% \footnote{\url{https://youtube.com/audiolibrary}}
\vspace{-0.1in}
\subsection{Evaluation}
\label{sec:eval}
\vspace{-0.05in}

Two groups of noisy sets are used for evaluation. The first is obtained by processing the test-clean subset of Librispeech with our room simulator to mimic the same three microphone array configuration used during training. Separate sets of 2620 utterances are generated at two SNR levels each for speech-based and non-speech based noise, using held-out noise segments from the training set.

The second group used in this study mimics the voice-search use case, unlike Librispeech. Short queries ($<$10 seconds) were re-recorded in a living-room lab. Desired speech and noise were recorded separately using a four microphone array, which differs from the triangular array used during training. The first two microphones were spaced 7.1~cm apart on the top of the device while the third and fourth were on the front and side, respectively. 100 different queries were played through a speaker from 7 different positions 4~m from the microphone array at a height of 1.5~m and a volume of 40~dB over ambient room noise. Two types of noise, speech-based from a movie and non-speech environmental noise, were separately played through a loudspeaker from the same 7 locations. Speech and noise from different locations were mixed at five different SNR levels with roughly 6 seconds of noise before the start of the query. 

\begin{table}[ht]
  \centering
  \caption{WER comparisons with proposed Cleanformer on the LibriSpeech set with added reverberation and noise.}
  \footnotesize
  \label{tab:librispeech-non-speech}
  \begin{tabular}{l|c|cc|cc}
    \hline
    \multirow{2}{2cm}{\textbf{Noisy LibriSpeech}}  &
    \multirow{2}{3em}{\textbf{Reverb Only}} &
    \multicolumn{2}{c}{\textbf{Non-Speech}}& 
    \multicolumn{2}{c}{\textbf{Speech}}\\
    {} &  {} & \multicolumn{1}{c}{\textbf{-5\,dB}} &  \multicolumn{1}{c}{\textbf{5\,dB}}  & \multicolumn{1}{c}{\textbf{-5\,dB}} & \multicolumn{1}{c}{\textbf{5\,dB}} \\
    \hline
    \textbf{Baseline} &$\textbf{7.2}$ &$36.5$  & $14.0$  &  $65.3$ &  $28.0$  \\
    Ideal Beamformer & $\textbf{7.2}$ &$31.8$ &  $12.3$ &$59.4$ & $25.0$   \\ 
    Speech Cleaner & $9.7$ &$14.3$ &  $12.4$ & $24.0$ & $18.4$  \\
    Cleaner Mask & $7.6$ & $17.0$ & $10.1$  & $32.6$ & $17.7$ \\
    \hline
   % Cleanformer & $7.3$ & $14.4$ & $9.3$ & $20.4$ &  $14.1$  \\
    \textbf{Cleanformer} & $7.3$ & $12.7$ & $8.9$ & $18.8$ &  $13.5$  \\
    \quad-- Mask Scalar Model & $7.4$ & $14.3$ & $9.3$ & $20.9$ &  $14.0$  \\
    % \quad-- Bad Context Training & $7.3$ & $13.3$ & $9.0$ & $19.1$ &  $13.6$  \\
    \quad-- Bad Context Training & $7.4$ & $\textbf{12.5}$ & $\textbf{8.8}$ & $\textbf{17.7}$ &  $\textbf{13.3}$  \\
    \hline
  \end{tabular}
  \vspace{-0.15in}
\end{table}

\begin{table*}[ht]

 \centering
 \caption{WER comparisons with proposed Cleanformer on the re-recorded sets with different sized arrays. }
  \label{tab:vendor-comparison}
  \footnotesize
  \begin{tabular}{c|l|c|ccccc|ccccc}
    \hline
     \multirow{3}{*}{\textbf{\# Mics}} &\multirow{3}{4em}{\textbf{Algorithm}}  &\multirow{3}{4em}{\textbf{Reverb Only}} & \multicolumn{5}{c}{\textbf{Speech-Based}} & \multicolumn{5}{c}{\textbf{Environmental}} \\
    {} & {}& {} & \multicolumn{5}{c}{\textbf{SNR(dB)}} & \multicolumn{5}{c}{\textbf{SNR(dB)}} \\
    {} & {}& {} & {-12} & {-6} & {0} & {6} & {12} & {-12} & {-6} & {0} & {6} & {12}  \\
    \hline
    {}& \textbf{Baseline} & 3.9 &97.9 &89.6 &53.7 &19.2 &7.5 &52.8 &40.5 &24.4 &11.8 &6.3 \\
    \hline

    \multirow{6}{*}{2} &Ideal Beamformer& \textbf{3.8}  & 94.4  & 83.2  & 39.1  & 12.1  & 6.0  &48.0  & 35.5  & 18.1  & 8.1  & 5.1 \\
    { } &Speech Cleaner&  4.5 & 56.9 & 29.4 & 13.8 & 8.3 & 6.2 & 42.7  & 22.3  & 11.1  & 8.3  & 6.6 \\
    { } &Cleaner Mask& 3.9 & 84.7 & 55.7 & 20.4 & 8.7 & 4.8 & 40.4 & 21.3 & 9.5 & 6.3 & 4.7 \\
   % { } &\textbf{Cleanformer} & 4.0  & 55.6  & 26.3  & 9.3  & 5.5  & 4.7  & 35.4  & 18.4  & 9  & 5.9  & 4.5 \\

    { } &\textbf{Cleanformer}&4.1&45.6 & 17.8 & \textbf{7.5} & 5.3 & 4.5 & \textbf{27.4} & \textbf{14.3} & 7.7 & \textbf{5.1} & 4.4\\
    { } &\quad-- Mask Scalar Model& 4.0  & 59.9  & 26.3  & 9.1  & 5.3  & 4.6  & 32.5  & 17.1  & 8.3  & 5.3  & 4.5 \\
    % { } &\quad-- Bad Context Training&3.9  & \textbf{45.2} & 19.0 & \textbf{8.0} & \textbf{5.0} & \textbf{4.3} & 29.8 & 16.1 & 8.1 & 5.3 & \textbf{4.4}\\
     { } &\quad-- Bad Context Training & 3.9 &  \textbf{38.5} &  \textbf{16.2} &  7.9 &  \textbf{5.0} &  \textbf{4.3} &  28.6 &  14.9 &  \textbf{7.0} &  \textbf{5.1} &  \textbf{4.3} \\
    \hline

    \multirow{6}{*}{3} &Ideal Beamformer& \textbf{3.8}  & 94.8  & 84.4  & 40.2  & 12.4  & 6.4  & 47.7  & 35.2  & 18.7  & 8.3  & 4.9 \\
    { } &Speech Cleaner&  4.8  & 22.9  & 10.8  & 7.7  & 6.5  & 6.0  & 21.1  & 10.5  & 8.3  & 6.8  & 6.0 \\
    { } &Cleaner Mask&  \textbf{3.8} & 71.5 & 36.0 & 13.7 & 5.6 & 4.3 & 26.2 & 12.0 & 6.9 & 5.3 & 4.4 \\
  %  { } &\textbf{Cleanformer} & 3.9  & 31.8  & 13.9  & 6.6  & 5.3  & 4.9  & 21.8  & 11.7  & 7.2  & 5.3  & 4.5 \\
    { } &\textbf{Cleanformer}&4.0&19.4&7.4&5.5&4.6&4.4&\textbf{12.2}&\textbf{6.7}&5.1&4.3&\textbf{4.2}\\
    { } &\quad-- Mask Scalar Model& 3.9  & 35.8  & 13.9  & 7.2  & 5.3  & 4.6  & 19.6  & 10.1  & 6.4  & 4.9  & 4.5 \\
    % { } &\quad-- Bad Context Training&4.0  & \textbf{19.2}  & 8.4  & \textbf{5.3}  & 4.8  & \textbf{4.3}  & 15.1  & 8.2  & 5.2  & \textbf{4.3}  & 4.4 \\
        { } &\quad-- Bad Context Training&3.9 &  \textbf{14.8} &  \textbf{7.3} &  \textbf{4.8} &  \textbf{4.5} &  \textbf{4.3} &  12.7 &  \textbf{6.7} &  \textbf{4.9} &  \textbf{4.2} &  4.3 \\
    \hline
    \multirow{6}{*}{4} &Ideal Beamformer& \textbf{3.7}  & 91.9  & 76.3  & 30.4  & 9.6  & 4.9  & 44.2  & 30.2  & 14.9  & 7.2  & 4.5\\
    { } &Speech Cleaner& 4.5  & 17.2  & 10.1  & 6.8  & 6.0  & 5.5  & 17.2  & 10.8  & 8.0  & 7.1  & 6.0  \\
    { } &Cleaner Mask&  4.0 & 71.3 & 36.5 & 13.9 & 6.1 & 4.6 & 24.0 & 11.4 & 7.0 & 5.0&	\textbf{4.3} \\
 %   { } &\textbf{Cleanformer} & 4.0  & 26.0  & 10.8  & 6.7  & 5.7  & 4.7  & 18.7  & 10.1  & 6.7  & 5.1  & 4.5 \\
    { } &\textbf{Cleanformer}& 4.0  & 14.9&7.2&5.2&4.7&4.6&\textbf{9.9}&6.1&4.9&4.5&\textbf{4.3}\\ 
    { } &\quad-- Mask Scalar Model&  3.9  & 30.8  & 11.1  & 6.1  & 5.2  & 4.6  & 15.4  & 9.3  & 6.2  & 4.6  & 4.5\\
    % { } &\quad-- Bad Context Training& 4.0  & \textbf{14.5}  & \textbf{7.4}  & \textbf{5.3}  & 4.8  &\textbf{4.5} & 11.7  & 6.4  & 5.0  & \textbf{4.0}  & \textbf{4.3}\\ 
    { } &\quad-- Bad Context Training& 3.9 &  \textbf{10.9} &  \textbf{6.6} &  \textbf{4.8} &  \textbf{4.3} &  \textbf{4.4} &  10.5 &  \textbf{5.6} &  \textbf{4.5} &  \textbf{4.3} &  4.4 \\

    %  3 & { } &  & & & & & & & & & & &
    %  4 & { } & 4 & & & & & & & & & & & &
    % \hline
    % \multirow{3}{*}{\textbf{Speech Cleaner}}& 2 & & & & & & & & & & & &
    %  { } & 3 & & & & & & & & & & & &
    %  { } & 4 & & & & & & & & & & & &
    % \hline
    % \multirow{3}{*}{\textbf{Cleanformer}}& 2 & & & & & & & & & & & &
    %  { } & 3 & & & & & & & & & & & &
    %  { } & 4 & & & & & & & & & & & & 
     
    %       $12$ & $7.4$ & $4.2$ & $4.4 $ & $4.5 $  \\
    %  $6$ & $18.9$ & $5.2$ & $4.9$ & $4.8 $  \\
    %  $0$ & $53.4$ & $10.8$ & $7.3$ & $6.6 $  \\
    %  $-6$ & $89.5$ & $32.3$ & $18.1$ & $14.6 $ \\
    %  $-12$ & $97.7$ & $62.9$ & $40.4$ & $35.6 $  \\
     
    %  Reverb & $3.5$ & $3.5$ & $3.3$ & $3.4 $  \\
    %  $12$ & $6.3$ & $4.9$ & $4.5 $ & $4.4 $  \\
    %  $6$ & $11.6$ & $5.3$ & $4.6$ & $4.7$  \\
    %  $0$ & $24.1$ & $8.7$ & $5.4$ & $4.9$  \\
    %  $-6$ & $40.3$ & $18.6$ & $9.9$ & $8.5$  \\
    %  $-12$ & $52.7$ & $34.9$ & $20.5$ & $17.0$  \\

    \hline
  \end{tabular}
 
\end{table*}

\vspace{-0.05in}
\section{Results}
\label{sec:results}

\begin{table}[ht]
  \centering
  \caption{WER for the noisy Librispeech set when the desired speaker is present in the noise context.}
  \label{tab:corrupt context}
  \footnotesize
  \begin{tabular}{l|cc|cc}
    \hline
    \multirow{2}{*}{\textbf{Noisy LibriSpeech}}  & 
    \multicolumn{2}{c}{\textbf{Non-Speech}}& \multicolumn{2}{c}{\textbf{Speech}}\\
    \multirow{2}{*}{\textbf{}}  &
    \multicolumn{1}{c}{\textbf{-5dB}}&
    \multicolumn{1}{c}{\textbf{5dB}}&
    \multicolumn{1}{c}{\textbf{-5dB}}&
    \multicolumn{1}{c}{\textbf{5dB}}\\

    \hline
    \textbf{Baseline}  & $36.5$ &   $14.0$ & $65.3$ &  $28.0$  \\
    Speech Cleaner & $95.5$ &  $99.0$ & $94.3$ & $98.7$ \\
    % Cleaned Mask & $51.1$ &  $25.2$ & $71.9$ &  $52.0$ \\
    \hline
   % Cleanformer & $88.7$&  $92.5$ & $90.0$ &  $92.3$  \\  
%    Cleanformer$++$ & $29.5$ &  $12.4$ & $57.0$ &  $19.5$  \\

    \textbf{Cleanformer} & $\textbf{28.8}$ &  $\textbf{12.5}$ & $\textbf{55.9}$ &  $\textbf{18.7}$  \\
    % \quad-- Bad Context Training & $89.8$ &  $93.8$ & $91.2$ &  $94.5$  \\
    \quad-- Bad Context Training & $98.3$ &  $95.7$ & $98.4$ &  $98.4$  \\
    \hline
  \end{tabular}
\vspace{-0.22in}
\end{table}

% \begin{figure*}[bt]
%   \centering
%   \includegraphics[scale=0.17]{noisy_60-123286-0003.w.pdf}
%   \caption{{Log-mel spectrograms for different stages of processing along with the estimated masks from the Cleanformer for three different SNR conditions.}}
%   \label{fig:spectrograms}
%   \vspace{-0.15in}
% \end{figure*}

For comparison, we present results using a few alternative techniques. ``Baseline'' uses just our ASR model with no enhancement frontend. ``Speech Cleaner'' shows the case where the single channel enhanced output is fed directly into the ASR model. ``Cleaner Mask'' is a mask formed by taking the ratio of the Speech Cleaner output and the noisy input. This mask is scaled and floored using $\alpha=0.5$ and $\beta=0.01$ to reduce distortion; otherwise, the masked output would be equivalent to that of Speech Cleaner:
\vspace{-0.1in}
\begin{equation}
\label{eq:masking}
\mathbf{CleanerMask}(t,f) =  \max(\left(\frac{{\mathbf {SpeechCleaner}}(n,f)}{{\mathbf Y}(n,f)}\right)^\alpha, \beta).
\end{equation} 
The purpose of this is to demonstrate that Cleanformer is performing more of a function than simply computing a mask directly from its two inputs. The final comparison is an \emph{ideal} beamformer output fed directly into the ASR model. A common technique steers the beamformer using  statistics of desired speech and noise estimated using time frequency masks obtained by a neural network \cite{higuchi2016robust,heymann2016neural,heymann2018performance}. Here, as an upper bound on performance, rather than being estimated through masking, the desired speech and noise statistics are \emph{directly} used to specify beamformer coefficients via a principal eigenvector steering mechanism \cite{higuchi2016robust}. 

Table \ref{tab:librispeech-non-speech} presents results using the simulated noisy Librispeech-based sets with separately added speech noise and non-speech environmental noise with the same three microphone array configuration used during training. Two SNR values are explored along with a reverberation-only condition with no added noise. 

In Table \ref{tab:vendor-comparison}, results are shown for the internal re-recorded datasets. Here the results are presented for two, three, and four microphone arrays as described in \ref{sec:eval}. It is important to note although the array size was varied from two to four microphones, the underlying Cleanformer model does not need to be aware of the number of microphones used or of the array geometry. It still receives one channel of raw input and one channel of enhanced input from the Cleaner. The model was only trained using the three channel triangular array data described in \ref{sec:training} and was not retrained for these cases. The only adjustment is the number of channels of input that the Speech Cleaner receives. 

Results from two variations of Cleanformer are also presented. ``-- Mask Scalar Model'' denotes Cleanformer using a fixed masked scalar value of $\alpha=0.5$ rather than having it selected by a model on a frame-wise basis. This $\alpha$ value was selected by tuning on development sets. For ``-- Bad Context Training", Cleanformer training was done without any of the noise contexts being replaced with either the desired speaker or white noise.

The trends are similar across both tables. Cleanformer significantly reduces WERs, especially in the lower SNR cases, while maintaining the WER in the clean cases. Table \ref{tab:vendor-comparison} shows that WERs improved with the number of microphones. In particular for \mbox{-12~dB} with speech noise, we see a relative error rate improvement of 53\% with 2 mics going to 85\% at 4 mics, with the WER going from 97.9 to 14.9. Better WERs are obtained using ``-- Bad Context Training'', but as we will show in Table~\ref{tab:corrupt context}, Cleanformer significantly outperforms it when the desired speaker is in the noise context.

The ideal beamformer also yielded gains that increased with the number of microphones but that were well below those of Cleanformer. Speech Cleaner yielded similar large gains as Cleanformer at low SNRs but degraded in clean cases. Cleaner Mask dulls the impact of the Speech Cleaner performing worse in lower SNRs but improves at higher SNRs due to reduced distortion.

Note that Speech Cleaner, Cleanformer and the ASR model are all causal, streaming models that do not use any right context, making the entire system well suited for streaming applications. To quantify this, we compute the overall ASR endpointer latency, defined as the time it takes for the ASR model to detect end-of-speech after the user stops speaking. Compared to the baseline of no enhancement frontend, when using Cleanformer, the ASR model detects end-of-speech roughly at the same time in clean conditions (2~ms faster, on average), 139~ms faster at 12~dB non-speech noise and 276~ms faster at 12~dB speech-noise. In noisy conditions, using Cleanformer enables the ASR model to detect end-of-speech sooner since it is less affected by background noise.

For the results discussed thus far, we made the assumption that the desired speaker was not present in the noise context. While this assumption holds in the majority of the use cases, its failure is still an important edge-case. Table \ref{tab:corrupt context} examines the impact of failing to meet this assumption for the noisy Librispeech-based sets where the noise context has been replaced with the query to ensure that the desired speaker is present. Recall the context is used only in the adaptation of the Speech Cleaner and is not passed on to a model. 
 The Speech Cleaner which will erase much of the desired speaker because it was present in the noise context causing deletions and a catastrophic error rate. ``-- Bad Context Training" which was not trained with any utterances where the desired speaker was in the noise context exhibits similar failures. In contrast, Cleanformer, while not attaining the performance improvement seen under the intended conditions, still show improvements ranging from 11\% to 30\% relative WER improvement. The model has likely learned to suppress background noise without fully relying on Speech Cleaner, making best possible use of the available partial information.

\vspace{-0.05in}
%\input{5.corrupt}
%\vspace{-0.05in}
\vspace{-1em}
\section{Conclusion}
\label{sec:conclusion}

This work introduced Cleanfomer \textemdash a streaming, array configuration-invariant neural enhancement frontend model for ASR. 
%As inputs, it takes a single channel each of raw and enhanced features. We obtained relative WER improvements often greater than 50\% across SNR levels for simulated and re-recorded data sets in both speech-based and non-speech based noise. 
Cleanformer, which takes a single channel of raw input and a single channel of enhanced input, showed relative WER improvements often greater than 50\% across SNR levels for simulated and re-recorded data sets in both speech-based and non-speech based noise. 
Improvement increased with the number of microphones in the array. There was also no adverse impact in clean conditions. The model was designed with the assumption that the desired speaker is not present in the noise context, which is expected in the majority of cases. We also demonstrate that training the model with a small percentage of utterances where the noise context is not as expected allowed Cleanformer to still achieve some gains when noise context contains the desired speaker. The model can be used without regard to the array size or configuration and does not need to be retrained for different arrays. Cleanformer represents a promising alternative architectural direction for combining signal processing and machine learning, demonstrating better applicability to streaming applications, like in smart speakers, than the commonly used mask-steered beamformer. Future work will focus on extending Cleanformer to more general use cases, like meeting transcriptions and long-form audio where the noise profile is expected to change during the ASR session.

% \section{ACKNOWLEDGMENTS}
% \label{sec:ack}
% The authors are grateful to Arden Huang for his early work on Hotword Cleaner. We also thank Nathan Howard, Sankaran Panchapagesan, Alex Park, James Walker, and Alex Gruenstein for helpful discussions and Andrew Sutter, Adam Whiteside, Frances Kwee, and Lawrence Lin for data collection help.


\bibliographystyle{IEEEbib}
\bibliography{refs}

\end{document}